# Electrodynamics of a ring-shaped spiral resonator

N. Maleeva,[1] M. V. Fistul,[1,2] A. Karpov,[1] A. P. Zhuravel,[3] A. Averkin,[1] P. Jung,[4] and A. V. Ustinov[1,4]

[1] *Laboratory of superconducting metamaterials, National University of Science and Technology "MISIS", Moscow, 119049, Russia*

[2] *Theoretische Physik III, Ruhr-Universität Bochum, Bochum, 44801 Germany*

[3] *B. Verkin Institute of Low Temperature Physics and Engineering, National Academy of Sciences of Ukraine, Kharkov, 61103, Ukraine*

[4] *Physikalisches Institut, Karlsruhe Institute of Technology (KIT), Karlsruhe, 76131, Germany*

We present analytical, numerical and experimental investigations of electromagnetic resonant modes of a compact monofilar Archimedean spiral resonator shaped in a ring, with no central part. Planar spiral resonators are interesting as components of metamaterials for their compact deep-subwavelength size. Such resonators couple primarily to the magnetic field component of the incident electromagnetic wave, offering properties suitable for magnetic meta-atoms. Surprisingly, the relative frequencies of the resonant modes follow the sequence of the odd numbers as $f_1:f_2:f_3:f_4\ldots=1:3:5:7\ldots$, despite the nearly identical boundary conditions for electromagnetic fields at the extremities of the resonator. In order to explain the observed spectrum of resonant modes, we show that the current distribution inside the spiral satisfies a particular Carleman type singular integral equation. By solving this equation, we obtain a set of resonant frequencies. The analytically calculated resonance frequencies and the current distributions are in good agreement with experimental data and the results of numerical simulations. By using low-temperature laser scanning microscopy of a superconducting spiral resonator, we compare the experimentally visualized ac current distributions over the spiral with the calculated ones. Theory and experiment agree well with each other. Our analytical model allows for calculation of a detailed three-dimensional magnetic field structure of the resonators.

## I. INTRODUCTION

Metamaterial is an artificially tailored media showing unusual electrodynamic properties based on the use of the resonant elements (Ref. 1). At certain frequencies, the effective magnetic permeability and the electrical permittivity both become negative, resulting in e.g. "left-handed" wave propagation and, therefore, a negative index of refraction. The standard design of such macroscopic structures, proposed over decade ago, is a set of resonators incorporating patterned metallic layers (Ref. 2, 3). The electrical and magnetic meta-atoms are designed as sub wavelength size micro resonators that couple primarily to either the electric or the magnetic field of the incoming electromagnetic wave. In first experiments, various types of split-ring resonators (SRR) have been used as magnetic meta-atoms in order to achieve effective negative magnetic permeability (Ref. 3-5). The resonance frequencies of such resonators are determined by the ratio between the width of the gap $l$ and the size $R$ of the SRR somewhat limiting the minimum size of the resonators. The planar spiral resonators (PSR) were proposed in order to drastically reduce the resonator size relative to the wavelength (Ref. 5, 6).

The PSR operates as a distributed resonator having multiple resonance modes and initial experiments with superconducting PSRs have shown numerous resonances (Ref. 6, 7). For example, a monofilar PSRs with densely packed turns within a diameter of 6 mm had the fundamental resonance mode at a frequency of 74 MHz. This corresponds to a ratio of the wavelength λ to resonator diameter $D$ of about 680 (Ref. 7), thus demonstrating the potential of spiral structures for designing ultra-compact resonators. Analyzing the experimental data reported in (Ref. 7), one can notice that the resonant frequencies depend on the mode number $n$ as $f_n \approx f_1 (2n - 1)$, thus following the odd number sequence with $f_1:f_2:f_3:f_4… \approx 1:3:5:7…$. This seems rather surprising fact, as the RF current flowing in the resonator has open boundary conditions on both extremities of the spiral. In a conventional one-dimensional transmission line cavity with two identical open ends, the spectrum of the modes is just a set of evenly spaced frequencies $f_1:f_2:f_3:f_4… \approx 1:2:3:4…$ (see Fig. 1).

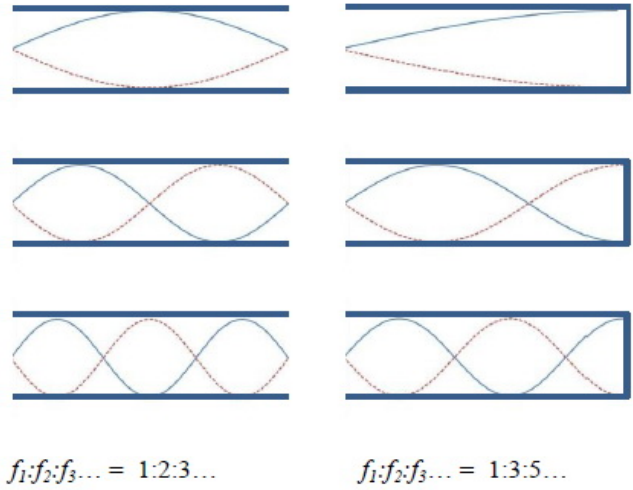

FIG. 1. The diagram showing the standing wave patterns for current in low-loss transmission lines with symmetric (left) and asymmetric (right) boundary conditions. The resonator with symmetric loads (open ends) has a spectrum of the resonant λ/2 modes with frequencies rising as integer numbers $f_1:f_2:f_3.. = 1:2:3...$ Uneven loads, like open circuit and short, lead to odd number ratio in the frequency spectrum of the resonances: $f_1:f_2:f_3... = 1:3:5...$ As we consider the spiral ring resonator (Fig. 2 a) with open circuit loads at the both extremities of the line, one may expect the resonance frequencies rising as integer numbers 1:2:3…. Contrary to the expectations, as one can see in Fig. 2, the resonance spectrum of the ring follows the set of odd numbers 1:3:5…, causing an intriguing question about the inner modes structure of this resonator.

As we turn to the theoretical description of the electrodynamic properties of a PSR, we notice that a very low resonance frequency in PSRs can be explained qualitatively by a simple model considering the spiral as one-dimensional transmission line resonator with the length $L$ rolled in $N$ turns. As both ends of the transmission line are open-ended, one can expect the first half-wave resonance at the frequency

$$f_1 = \frac{c}{2L},$$

where $c$ is the speed of light, and the $L$ is the length of the transmission line. Assuming that for a narrow monofilar Archimedean spiral with no central part, all the turns have approximately the same radius $R$:

$$L \approx 2\pi RN, \qquad (1)$$

with $N$ being the number of turns, one can obtain the $n$ half-wave mode resonance frequency $f_n$ as

$$f_1 \approx \frac{c}{4\pi RN} \text{ and } f_n \approx \frac{nc}{4\pi RN} \qquad (2)$$

where $n$ = 1, 2 3, … Thus, for the fundamental mode $n$ = 1 of a resonator with, e.g., 100 turns, one expects the reduction of the diameter to wavelength ratio by two orders of magnitude. The ease to create a deep sub-wavelength, compact resonator, using a planar spiral, is making this resonator a natural candidate for the magnetic component in metamaterials.

The rough estimate (2), as shown below, explains the first resonance frequency within a factor of two and gives a good idea of compactness of the resonator. Nevertheless, this simplified approach fails to explain the spectrum of higher modes. The spiral resonator has open circuit boundary conditions at both extremities of its conductive line and the expected spectrum has components evenly spaced in frequency as in the left part of Fig. 1. Nevertheless, experiments (Ref. 6, 7) have shown that spectrum follows the odd number series as in the right part of Fig. 1, indicating a deficiency of the simplified approach. The difficulty to interpret the experimental data (Ref. 6, 7) for ring-shaped monofilar Archimedean resonators in a simple qualitative way makes it important to develop an analytical model in order to clarify the origin of the odd series spectrum and the current distribution of the standing waves at the resonance frequencies.

In this paper, we present detailed experimental characteristics of a ring-shaped monofilar Archimedean RF resonator, develop the analytical model of the resonator, and finally use the obtained analytical relations to explain the modal structure and spectrum observed in experiment. All experiments have been carried out with planar Nb superconducting spiral resonators.

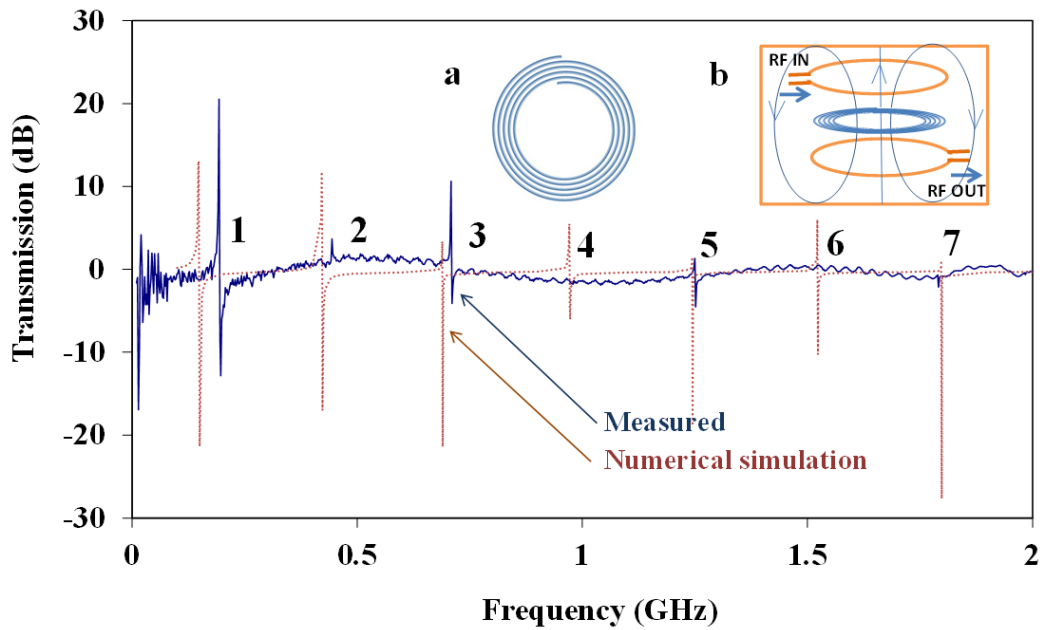

FIG. 2. Experimental and numerically simulated RF response of a ring-shaped superconducting monofilar Archimedean resonator. The numbers at the peaks denote the mode number $n$. The resonator is a superconducting 40-turn Nb spiral with external diameter of 3 mm and internal diameter of 2.2 mm. The spiral is deposited on quartz substrate. A simplified view of the spiral (with less turns) is shown in insert (a). The data are obtained by measuring transmission magnitude $|S_{21}|$ through a spiral resonator in a setting shown in insert (b). The superconducting spiral was placed inside the cryostat and weakly magnetically coupled to a pair of loops, one

above and another below the spiral. The transmission coefficient $S_{21}$ was measured twice: with the resonator material in the normal state at 10 K ($S_{21}$ 10 K), and after cooling the sample to the temperature of liquid helium 4.2 K, ($S_{21}$ 4 K), at which Nb is superconducting. The data plotted by the blue solid line are the ratio of transmission $|S_{21}$ 4k$|$ / $|S_{21}$ 10 K$|$. The measured response of a weakly coupled resonator represents a series of narrow peaks, located at the resonance frequencies of the spiral resonator. The ratio of the resonance frequencies of a ring-shaped spiral resonator closely follows the set of odd numbers: $f_1:f_2:f_3:f_4...=1:3:5:7...$ The higher order resonance peaks are rapidly getting weaker, and the even peaks are generally weaker than the odd ones. The numerical simulations (dotted line) are in reasonably good agreement with the experimental results. The simulated $S_{21}$ curve is obtained as a difference in simulated transmission data for the superconductive (ideal conductor) and for a lossy resonator material of the same geometry, similar to the experiment.

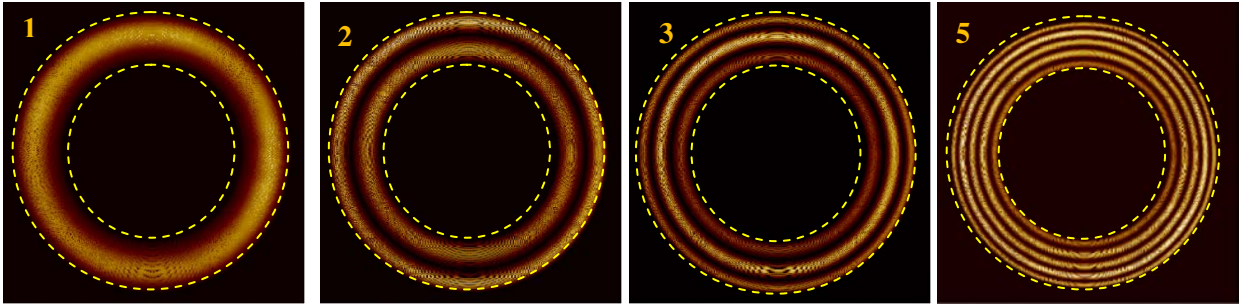

FIG. 3. Scanning laser-microscope images of the standing waves at the resonance frequencies of a ring-shaped spiral niobium resonator. The numbers at the images denote the mode number $n$ and the dashed line marks the boundaries of the Archimedean spiral. The laser beam scanned across the sample is locally suppressing superconductivity in the resonator, affecting the transmission of RF signal ($S_{21}$) through the resonator (measured in experimental setting shown in Fig. 1b). The contrast is proportional to the variation of the transmission coefficient $S_{21}$, which is recorded and plotted as a function of the laser beam position, thus mapping the amplitude distribution of RF current in the superconducting resonator. The strongest response (brighter regions) is located in the areas with the maximum RF current. One can note that the standing waves in the spiral resonator show a high symmetry, and that the number of RF current antinodes along the radial direction corresponds to the resonance number *n*.

## II.  EXPERIMENT

A typical example of measured spectral response of a weakly coupled superconducting Nb monofilar resonator is presented in Fig. 2. The spiral has the shape of a ring, with no central part (Fig. 2a). The resonator structure is etched out of a 200 nm thick Nb film deposited at on a quartz substrate. The width of the resonator conductor line is 5 μm, the spiral step is 10 μm, the outer diameter is 3 mm, and the number of turns is 40. Following Ref. 7, measurements were done in transmission mode by using two loops terminating the transmitting (Fig. 2b "in") and receiving (Fig. 2b "out") ports of the coaxial cables connected to a vector network analyzer. The sample was mounted in a liquid helium cryostat (Fig. 2b). At the resonance frequency, the resonator provides a stronger coupling between the loops, giving a narrow spike in the transmission versus frequency dependence curve $S_{21}(f)$. In order to calibrate our setup, we measured the transmission $S_{21}(f)$ through the system at a temperature above the critical temperature of Nb, close to 10 K ($S_{21}$ 10 K). At this temperature, the resonator material (Nb) is in the normal state, and, due to the losses in Nb, the resonator has nearly no effect on the transmission $S_{21}(f)$. The data plotted in the Fig. 2 are the measured transmission at 4.2 K (with superconducting Nb, $S_{21}$ 4.2K), normalized with respect to the measured transmission spectrum at 10 K (with normal Nb, $S_{21}$ 10 K). One can see sharp peaks at the resonance frequencies with the contrast of up to 20 dB. The frequencies of resonances approximately follow the odd numbers series: $f_1:f_2:f_3:f_4… ≈ 1:3:5:7…$ The amplitude of the response drops rapidly with increasing of the mode number, indicating a weaker coupling at higher modes.

The standing wave patterns of a similar spiral resonator have been tested with the low-temperature laser scanning microscope (Ref. 7). Experimental images shown in Fig. 3 are presenting the current amplitudes of standing waves at the resonance frequencies of a spiral resonator of the shape shown in Fig. 2a. The numbers at the images denote the resonance mode number $n$ and the dashed lines indicate the spiral boundaries. A laser beam focused on the surface of the superconductor leads to a local increase of temperature and to breaking of Cooper pairs via direct absorption of photons. Both these effects lead to a local depression of superconductivity in the Nb circuit, affecting the transmission of the RF signal ($S_{21}$) through the resonator in the experimental setup presented in Fig. 2b. The variation of $S_{21}$ is recorded and plotted as a map as a function of the laser beam position on the sample. The strongest response (brighter area) occurs in the parts of the resonator with the largest RF current amplitude, which has the major contribution to the transmitted signal. One can note that the standing waves in the spiral resonator have nearly perfect cylindrical symmetry, and that the number of maxima in the radial direction corresponds to the mode number $n$ of half-wavelengths $\lambda/2$ fitting the spiral extremities.

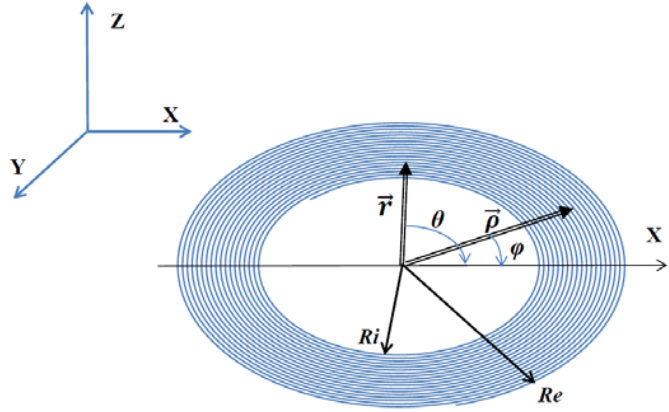

FIG. 4: Sketch of a ring-shaped spiral resonator with many turns. The polar coordinates $\rho$ and $\varphi$ are the coordinates of the point on spiral; $r$ and $\theta$ are the coordinates of the observation point in the plane of the spiral. $R_e$ and $R_i$ are external and internal radiuses of spiral, accordingly.

## III. ELECTRODYNAMIC MODEL OF A PLANAR SPIRAL RESONATOR

In the following we present an analytical approach to study the electrodynamics of planar spiral resonators. Our goal is to obtain analytical expressions for the frequencies of resonant modes, as well as electric and magnetic field spatial distributions around the spiral resonator. We study a ring-shaped monofilar spiral Archimedean resonator of a finite length and with many densely packed turns. The schematic of such a spiral resonator is shown in Fig. 4. We consider the case of lossless conductor, which corresponds to a superconducting spiral. The total number of turns is $N$. The current $\vec{j}(\vec{\rho},t)$ flows along the spiral. Such a geometry is characterized by the angle $\varphi$ which changes from 0 to $2\pi N$, and a corresponding change of a polar coordinate $\rho$. The equation of an Archimedean spiral is written as

$$\rho(\varphi) = R_e(1-\alpha\varphi), \qquad (3)$$

where the parameter $\alpha = \dfrac{d}{2\pi R_e} = \dfrac{R_e - R_i}{2\pi R_e N} \ll 1$, $R_e$ and $R_i$ are the external and the internal radius of the spiral, and $d$ is the distance between the adjacent turns.

The following analysis is similar to the one carried out in Refs. 8, 9 for an infinite helical coil. We neglect the transversal current inhomogeneity in the wire, and present the coordinate and the time dependent vector-potential in the following form (in cylindrical coordinates):

$$\vec{A}(z,r,\theta) = e^{i\omega t}\frac{\mu_0 I}{4\pi}\int \frac{e^{-ikR}}{R}\psi(s)d\vec{s}, \qquad (4)$$

where $I$ is the amplitude of the current excited in the spiral, $k = \dfrac{\omega}{c}$ is the wave vector, $s$ is the coordinate along the spiral, and $\psi(s)$ describes normalized inhomogeneous current flowing along the spiral. Here $R$ is the distance between the observation point with coordinates $(z,r,\theta)$ and the point on the spiral with coordinates $(0,\rho,\varphi)$.

The electric $\vec{E}(\vec{r},t)$ and magnetic $\vec{H}(\vec{r},t)$ fields are determined via the vector-potential $\vec{A}$ as

$$\vec{E}(\vec{r},t) = \frac{1}{i\omega\varepsilon_0\mu_0}\vec{\nabla}(\vec{\nabla}\vec{A}) - i\omega\vec{A}, \qquad (5)$$

$$\vec{H}(\vec{r},t) = \frac{1}{\mu_0}\left[\vec{\nabla}\times\vec{A}\right]. \qquad (6)$$

In order to obtain the coordinate dependence of the vector-potential we use the following geometrical relationships:

$$ds_r = d\varphi\left[\rho\sin(\theta-\varphi) - R_e\alpha\cos(\theta-\varphi)\right], \qquad (7)$$

$$ds_\theta = d\varphi\left[\rho\cos(\theta-\varphi) - R_e\alpha\sin(\theta-\varphi)\right]. \qquad (8)$$

On other hand, the distance $R$ can be expressed as

$$R = \sqrt{z^2 + D^2} = \sqrt{z^2 + r^2 + R_e^2 (1-\alpha\varphi)^2 - 2rR_e (1-\alpha\varphi)\cos(\varphi-\theta)}, \qquad (9)$$

We use the well-known representation of $e^{-ikR}/R$ in the following form (Ref. 10):

$$\frac{e^{-ikR}}{R} = \int_0^\infty \frac{x\,dx}{4\pi\sqrt{x^2-k^2}} J_0(Dx) e^{-\sqrt{x^2-k^2}|z|}. \qquad (10)$$

From Eq. (10) one can see that the electromagnetic field strongly decays in the direction perpendicular to the spiral plane. Using the relationship (Ref. 10)

$$J_0(Dx) = \sum_{m=-\infty}^{\infty} e^{im(\varphi-\theta)} J_m(xr) J_m\left[xR_e(1-\alpha\varphi)\right] \qquad (11)$$

we obtain the vector-potential $A(z; r;\varphi)$ as

$$A_r = e^{i\omega t} \frac{\mu_0 I}{4\pi} \sum_{m=-\infty}^{\infty} \int_0^{2\pi N} d\varphi \psi(s) \frac{ds_r}{d\varphi} e^{im(\varphi-\theta)} \int_0^\infty\int_0^\infty \frac{xe^{-\sqrt{x^2-k^2}|z|} dx}{4\pi\sqrt{x^2-k^2}} J_m(xr) J_m\left[xR_e(1-\alpha\varphi)\right] \qquad (12)$$

and

$$A_\theta = e^{i\omega t} \frac{\mu_0 I}{4\pi} \sum_{m=-\infty}^{\infty} \int_0^{2\pi N} d\varphi \psi(s) \frac{ds_\theta}{d\varphi} e^{im(\varphi-\theta)} \int_0^\infty\int_0^\infty \frac{xe^{-\sqrt{x^2-k^2}|z|} dx}{4\pi\sqrt{x^2-k^2}} J_m(xr) J_m\left[xR_e(1-\alpha\varphi)\right]. \qquad (13)$$

Taking into account that the most important contribution occurs from the terms with $m=\pm 1$, we simplify (13) and (14) as:

$$A_r = e^{i\omega t} \frac{\mu_0 I}{4\pi} \int_0^\infty d\rho \psi(\rho) \int_0^\infty \frac{xe^{-\sqrt{x^2-k^2}|z|} dx}{4\pi\sqrt{x^2-k^2}} J_1(xr) J_1(x\rho), \qquad (14)$$

$$A_\theta = e^{i\omega t} \frac{\mu_0 I}{4\pi} \int_0^\infty d\rho \frac{\rho\psi(\rho)}{R_e \alpha} \int_0^\infty \frac{xe^{-\sqrt{x^2-k^2}|z|} dx}{4\pi\sqrt{x^2-k^2}} J_1(xr) J_1(x\rho). \qquad (15)$$

In Eqs. (14) and (15), the function $\psi(\rho)$ describes the *inhomogeneous current distribution across the spiral*. In order to obtain the resonant frequencies we apply boundary conditions specific to the spiral case, i.e. the component of electric field parallel to the wire $E_s$ is equal zero. This condition can be written as

$$R_e \alpha E_r + r E_\theta \big|_{z=0} = 0. \qquad (16)$$

Using (5) and (6) we obtain

$$E_r = \frac{1}{i\omega\varepsilon_0\mu_0} \frac{d}{dr}\left[\frac{1}{r}\frac{d}{dr}(rA_r)\right] \qquad (17)$$

and

$$E_\theta = -i\omega A_\theta. \qquad (18)$$

To obtain the resonant frequencies, we use the following approximations: The wave vector $k$ is much smaller than a typical inverse size of inhomogeneities in current distribution $\psi(\rho)$, i.e. $k \ll \frac{1}{R_e - R_i} \ll \frac{1}{R_e}$, and the planar spiral is a rather narrow one, i.e. $R_e - R_i \ll R_e$.

Taking into account all the previous, Eq. (17) can be rewritten as

$$E_r = \frac{1}{i\omega\varepsilon_0\mu_0} \frac{d^2}{dr^2} A_r$$

In this case, Eqs. (15) and (16) can be also greatly simplified as

$$A_r(z=0) = \alpha A_\theta(z=0) = \frac{\mu_0 I e^{i\omega t}}{(4\pi)^2} \int_{R_i}^{R_e} d\rho \psi(\rho) \ln \frac{R_e}{|\rho - r|} . \qquad (20)$$

Moreover, by making use of Eqs. (17), (18) and (19) we obtain the dependence of the vector-potential $A_r$ at the coordinate $r$ (for $R_i < r < R_e$, i.e. inside the spiral) in a simple oscillatory form:

$$A_r(z=0) = \frac{\mu_0 I e^{i\omega t}}{R_e (4\pi)^2} F(r) \qquad (21)$$

$$F(r) = \cos(pr + \phi), \qquad (22)$$

where the wave vector $p$ is determined as $p = \omega/(c\alpha)$.

The parameters $p$ and $\phi$ and a corresponding form of the current distribution have to be found from the integral equation:

$$\int_{R_i}^{R_e} d\rho \, \psi(\rho) \ln \frac{R_e}{|\rho - r|} = F(r). \qquad (23)$$

The solution of this so-called Carleman type singular integral equation is written in the following form (Ref. 11):

$$\psi(x) = \frac{1}{\sqrt{1-x^2}} \left[ \int_{-1}^{1} \frac{\sqrt{1-y^2} F'(y) dy}{y - x} - \frac{1}{\ln \frac{2R_e}{w}} \int_{-1}^{1} \frac{F(y) dy}{\sqrt{1-y^2}} \right]. \qquad (24)$$

Here, $2w = R_e - R_i$ is the spiral's width, and the dimensionless coordinate $x$ is counted from the middle of spiral, i.e. $-1 < x < 1$. Substituting (22) into (24) we obtain the current distribution as:

$$\psi(x) = \frac{1}{\sqrt{1-x^2}} \left[ \int_{-1}^{1} \frac{\sqrt{1-y^2} \sin(pwy + \phi) dy}{y - x} + \frac{\pi J_0(pw) \cos(\varphi)}{\ln \frac{2R_e}{w}} \right], \qquad (25)$$

where $J_0(x)$ is the Bessel function of the first kind (Ref. 10).

The current distribution $\psi(x)$ has to satisfy the open-ended spiral boundary conditions $\psi(-1)=\psi(1)=0$. It is possible to satisfy these boundary conditions only for particular values of $p_n$ and therefore $\omega_n$. Moreover, there are even solutions for $\psi(x)$ as $\sin(\phi)=0$, and odd solutions as $\cos(\phi)=0$. For the even solutions, we obtain the transcendent equation determining the values of $p_n$ as:

$$\ln\frac{2R_e}{w} pwJ_1(pw) = J_0(pw). \qquad (26)$$

For the odd solutions, we obtain another transcendent equation as

$$J_0(pw) = 0. \qquad (27)$$

Finally, we obtain the set of resonant frequencies as

$$f_n = \frac{c\alpha p_n}{2\pi} = \frac{c\alpha}{4w}\gamma_n, \qquad (28)$$

where $\gamma_n = \frac{\pi}{2} p_n w$ can be found from Eqs. (26) and (27). The index $n$ corresponds to mode number and it is the number of oscillations of the current distribution $\psi(x)$ inside the spiral. Taking into account $\alpha = \frac{d}{2\pi R_e} = \frac{R_e - R_i}{2\pi R_e N} \ll 1$ and $2w = R_e - R_i$, we obtain:

$$f_n = \frac{c}{2L}\gamma_n = f_0 \gamma_n, \qquad (29)$$

here $n = 1, 2, 3 \ldots$, and $L$ is the total length of the conductive line in the spiral resonator, and $f_0 = c/2L$ coincides with the fundamental resonance frequency of a straight-line resonator of length $L$.

**Table I**     Resonance frequency correction factor

| Mode number $n$ | Correction factor $\gamma$ | Correction factor $2\gamma$ |
|---|---|---|
| 1 | 0,53 | 1,05 |
| 2 | 1,53 | 3,07 |
| 3 | 2,50 | 5,01 |
| 4 | 3,52 | 7,04 |
| 5 | 4,50 | 9,01 |
| 6 | 5,52 | 11,04 |
| 7 | 6,51 | 13,02 |
| 8 | 7,50 | 15,01 |

For parameters of the sample used in our experiment we obtained the frequency correction factor values $\gamma_n$ listed in Table I. Notice here, that the calculated sequence of the resonance frequencies for the modes of the ring-shaped spiral resonator is closely following the odd numbers series 1:3:5:7…, thus yielding an appropriate prediction of the spectrum of the inner modes. The fundamental resonance frequency $f_1$ of a ring-shaped spiral is very close to half of the resonance frequency (½ $f_0$) of a straight-line resonator with the same length of conductor. For the higher order resonances, $\gamma_n$ is asymptotically approaching the mode number $n$, giving a result similar to our simplified estimate (2). The resonance frequencies are equidistant along the frequency axis with an increment equal to $f_0 = c/2L$, similarly to the case of a straight-line resonator.

The above analytical model makes it possible to calculate the profile of the standing waves of the modes of the spiral using Eq. (25) and, therefore, to calculate the distribution of the magnetic field around the resonator. The predicted resonance frequencies and RF current distributions are compared with experimental data in the following section.

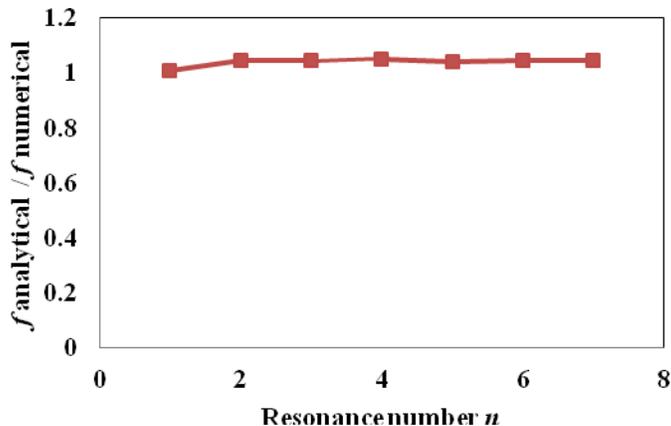

FIG. 5. Analytically and numerically calculated resonance frequencies appear nearly identical for the first seven modes. The systematic offset of about 4% can be related to a finite (non-zero) width of the resonator ring.

## IV.   RESULTS AND DISCUSSION

First, in order to verify our analytical prediction for the resonance mode frequencies, we performed numerical simulation with HFSS using the same spiral resonator geometry as in the analytical model. In both models we consider a free-standing spiral without substrate. The resonance frequencies predicted by numerical and analytical models match well, having a constant relative offset of about +4% (see Fig. 5). We suppose that this discrepancy occurs due to the fact that the spiral ring used in the simulation (and experiment) is not as narrow as expected in the analytical model, having an inner to outer diameter ratio of about 0.7. Remarkably, both analytical and numeric sets of resonance frequencies closely follow the sequence of the odd numbers $f_1:f_2:f_3...= 1:3:5...$.

Second, we compare the analytical model prediction with the experimental data. In the experiment (Fig. 2), the superconducting spiral resonator is located on a fused quartz substrate with a dielectric constant of about $e_r = 3.8$. It is known that the effective dielectric constant for a structure at the interface of the dielectric and air is approximately: $e_{eff} = (e_r +1)/2$. To explain our experiment with the analytical model we thus have to assume $e_{eff}=2.4$, which reduces the speed of light by the square root of the effective dielectric constant to about 0.65 of its value in vacuum. Experimental data and model prediction for the first seven modes are summarized in Table II. Most of resonance frequencies $f_{an}$ predicted by analytical model match experimental data $f_{exp}$ within 3%. We note that the most significant discrepancy between analysis and experiment is at the fundamental resonance frequency. We argue that this deviation is related to a strong coupling of the resonator fundamental mode to the input and output ports of our setup (wire loops that terminate coaxial lines near the sample), which is not accounted for by the model. The magnetic coupling strength changes for different modes due to the varying distribution of RF magnetic field around the spiral. A stronger resonator coupling leads to a bigger effect of the reactance of coupling elements, causing a shift in the resonator frequency.

**Table II.**      **Experimental and analytical resonance frequencies**

| $n$ | Experiment $f_{exp}$ (GHz) | Analytical $f_{an}$ (GHz) | $f_{an}/f_{exp}$ |
|---|---|---|---|
| 1 | 0.195 | 0.150 | 1.29 |
| 2 | 0.444 | 0.440 | 1.01 |
| 3 | 0.709 | 0.718 | 0.99 |
| 4 | 0.981 | 1.009 | 0.97 |
| 5 | 1.250 | 1.291 | 0.97 |
| 6 | - | 1.582 | - |
| 7 | 1.789 | 1.866 | 0.96 |

The waveforms of the inner modes of the spiral resonator are given by Eq. (25) of the analytical model. We calculated the normalized current amplitudes for the first three resonance modes. The results are presented in Fig. 6. In the same plot we also show the measured amplitudes for the same modes. The amplitude is plotted as a function of the distance along the radial direction. The abscise coordinate -1 corresponds to the inner edge of the ring-shaped spiral, and the coordinate 1 is at the outer edge of the spiral, as in Eq. (25). The experimental values of the amplitudes are obtained from the scanning laser microscopy data presented earlier in Fig. 3. The measured laser beam photo-response signal (Ref. 13) has a profile proportional to the local value of RF current density squared, $J_{RF}^2(x,y)$. One can see that the model is predicting the location of the maxima and minima rather well in the standing wave structure for all three modes.

Figure 7 presents a comparison between the measured and calculated current amplitudes for the 5$^{th}$ mode of the spiral resonator. The agreement of prediction with experiment is very good, with exception of the data at the inner edge of the spiral, apparently related to some experimental imperfection. We suppose that shorter lateral lobes in the standing wave structure adjacent to the edges of the spiral can be related to a higher phase velocity of electromagnetic waves at the edges of the spiral. The same feature can be also seen in Fig. 6. It can be argued that the enhanced phase velocity of light at the inner and outer extremities of the spiral leads to the shift of the series of resonant mode frequencies towards the series of odd numbers $f_1:f_2:f_3\ldots=$ 1:3:5…, which is different from the case of open-ended straight-line resonator characterized by evenly spaced frequencies $f_1:f_2:f_3\ldots=$ 1:2:3…(or, equivalently, $f_1:f_2:f_3\ldots=$2:4:6…).

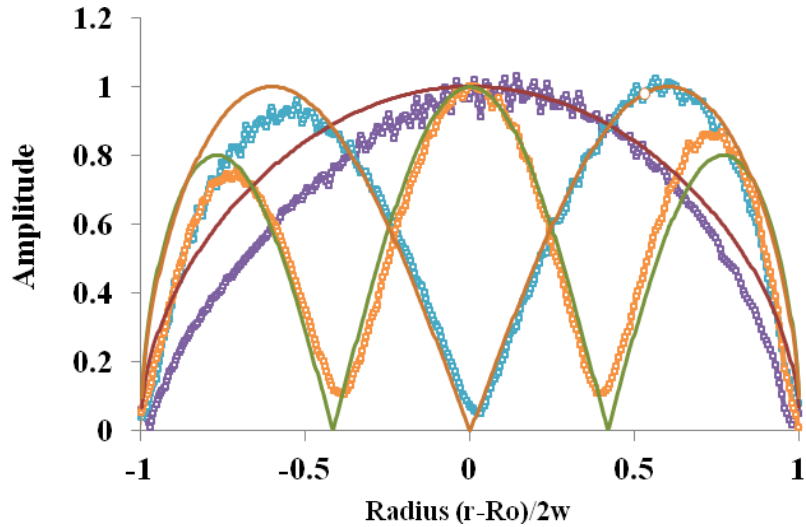

FIG. 6. Measured and analytically calculated amplitudes of the first three resonance modes of the ring-shaped spiral resonator. RF-current amplitude is plotted as a function of radial coordinate across the spiral width. The coordinate -1 corresponds to the inner edge of the ring-shaped spiral and the coordinate 1 is at the outer edge of the spiral. The experimental data are extracted from measurements presented in Fig. 3. The inner to outer spiral radius ratio is about 0.7. The model shows very good agreement with experimental results, even for a not so narrow spiral.

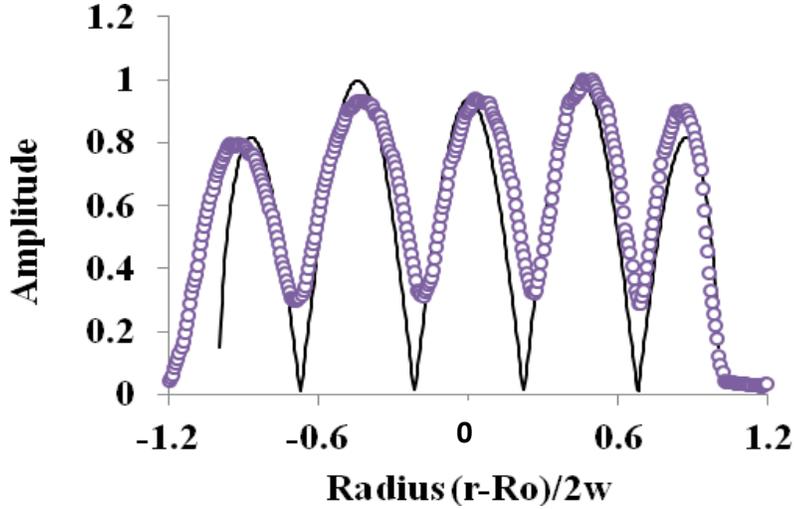

FIG. 7. Measured and calculated amplitudes of the 5$^{th}$ resonance mode of the ring-shaped spiral resonator. The experimental data are extracted from measurements presented in Fig. 3. The analytical model agrees well with experiment, predicting the locations of the minima and the shape of the standing RF-current wave. Note the shorter lateral lobes in the standing wave structure adjacent to the edges of the spiral. They indicate a higher phase speed of waves at the edges of the spiral.

     We used Eq. (25) in order to calculate the magnetic field structure around the resonator at the different modes. Magnetic field patterns are symmetrical with respect to the axis of symmetry of the spiral, and one can get complete information about the field structure by considering its distribution in the vertical plane containing the symmetry axis, as presented in Fig. 8. Here the vertical dotted line is the spiral axis of symmetry, and the horizontal straight line marks the spiral plane. The bold straight lines are the cross-sections of the ring-shaped spiral conductive strip and the digits are the mode numbers. Starting with the second mode, the magnetic field is mainly confined to the near field area of the resonator, about twice the diameter of the spiral. This feature leads to a reduction of the spiral coupling to the external magnetic RF

field for the higher modes, as already argued above. Another apparent attribute is that for the even modes the coupling is weaker than for the odd modes.

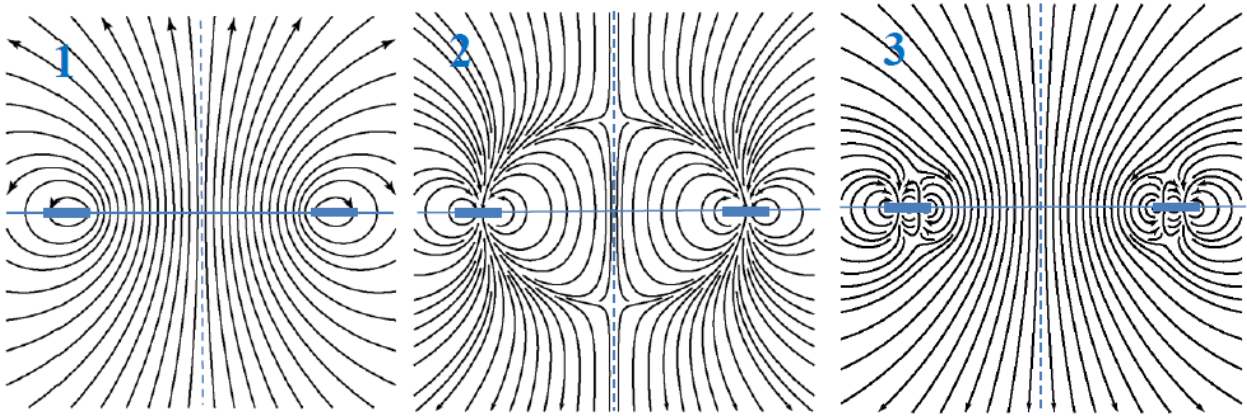

FIG. 8. Magnetic field lines calculated in a plane perpendicular to the spiral at the first three resonance frequencies. The vertical dotted line is the spiral axis of symmetry and the horizontal straight line marks the spiral plane. The bold straight lines are for the cross-sections of the ring-shaped spiral conductive strip and the digits stand for the corresponding mode numbers. Starting with the second mode, the magnetic field is mainly confined to the near field area, about twice the diameter of the spiral.

## V. CONCLUSION

We developed an analytical model quantitatively describing the electrodynamics of superconducting monofilar Archimedean spiral resonators. The studied planar spiral resonator is shaped as a ring, with no central part. In order to explain the observed spectrum of the inner modes, we demonstrated that the current distribution inside of a spiral satisfies a particular Carleman type singular integral equation. By solving such an equation, we obtained a set of resonant frequencies. The calculated resonance frequencies and the current distributions are found to be in good agreement with experimental data and the results of numerical simulation. The analytical model derives detailed patterns of the RF magnetic field around the resonator and makes it possible to explain the strength of its coupling to the measurement circuitry. Compared to numerical simulation, the developed analytical model of a ring-shaped planar spiral resonator has a substantial advantage for clarifying the structure of inner modes and may substantially accelerate evaluation of the characteristics of metamaterial elements.

**ACKNOLEDGEMENTS**

We would like to acknowledge Steven M. Anlage for fruitful discussions and for providing us with a sample for the presented experiments. This work was supported in part by the Ministry of Education and Science of the Russian Federation, by the Russian Foundation of Basic Research, and also by the Deutsche Forschungsgemeinschaft (DFG) and the State of Baden-Württemberg through the DFG Center for Functional Nanostructures (CFN). Philipp Jung acknowledges the financial support from the Helmholtz International Research School for Teratronics (HIRST).

**REFERENCES**

1 N. Engheta, R. W. Ziolkowski, Metamaterials: Physics and Engineering Explorations , Wiley & Sons, New York, (2006); S. Zouhdi, Ari Sihvola, Al. P. Vinogradov, Metamaterials and Plasmonics: Fundamentals, Modelling, Applications. , Springer-Verlag, New York, (2008).

2 J. B. Pendry, A. J. Holden, D. J. Robbins, W. J. Stewart, Magnetism from conductors and enhanced nonlinear phenomena , IEEE Trans. Microwave Theory Tech, 47, 2075, (1999).

3 R. A. Shelby, D. R. Smith, S. C. Nemat-Nasser, S. Schultz, Microwave transmission through a twodimensional, isotropic, left-handed metamaterial, Appl. Phys. Lett., 78, 489 (2001); R. A. Shelby, D. R. Smith, S. Schultz, Experimental Verification of a Negative Index of Refraction, Science, 292, 77 (2001).

4 M. C. Ricci and S. M. Anlage, Single superconducting split-ring resonator electrodynamics, Appl. Phys. Lett. 88, 264102 (2006).

5 M. C. Ricci, N. Orlofi, S. M. Anlage, Superconducting metamaterials, Appl. Phys Lett. 87, 034102 (2005).

6 C. Kurter, J. Abrahams, S. M. Anlage, Miniaturized Superconducting Metamaterials for Radio Frequencies, Appl. Phys. Lett. 96, 253504 (2010).

7 C. Kurter, A. P. Zhuravel, J. Abrahams, C. L. Bennett, A. V. Ustinov, S. M. Anlage, Superconducting RF Metamaterials Made with Magnetically Active Planar Spirals,IEEE Trans. Appl. Supercond. 21, 709 (2011).

8 S. Kh. Kogan, Rasprostranenie voln vdol beskonechnoi spirali , DAN, 66, 867 (1949) [in Russian].

9 R. A. Silin and V. P. Sazonov, Zamedlyaushie Sistemi, Sov. Radio, Moscow (1966) [in Russian].

10 M. Abramovitz and I. A. Stegun, Handbook of Mathematical Functions with Formulas, Graphs, and Mathematical Tables, New York: Dover Publications, (1972).

11 A. D. Polyanin and A. V. Manzhirov, Handbook of Integral Equations, Second Edition, Chapman & Hall/CRC Press, Boca Raton, (2008).

12 L. D. Landau, E. M. Lifshitz, Quantum Mechanics: Non- Relativistic Theory, Elsevier, (1981).

13 A. P. Zhuravel, C. Kurter, A. V. Ustinov, and S. M. Anlage, Phys. Rev. B. 85, 134535 (2012).